\documentclass[a4paper,onecolumn,11pt]{quantumarticle}
\pdfoutput=1
\usepackage[utf8]{inputenc}
\usepackage[english]{babel}
\usepackage[T1]{fontenc}
\usepackage{amsmath}
\usepackage{amssymb} 
\usepackage{amsthm}
\usepackage{amsfonts}
\usepackage{hyperref}
\usepackage{booktabs}          % for nice-looking tables
\usepackage{tcolorbox}         % for colored boxes
\usepackage{multirow}
\usepackage{tabularx}
\usepackage[ruled, vlined, linesnumbered,nofillcomment]{algorithm2e}
\usepackage[strings]{underscore}
\usepackage{float} 
\usepackage{stmaryrd}
\usepackage{subcaption}
\usepackage{siunitx}
\usepackage{tikz}
\usetikzlibrary{positioning,calc,arrows.meta}
\usepackage{expl3}

\usepackage{lipsum}

\usepackage[numbers]{natbib}

\def\G{\mathcal{G}}

\newcommand{\wt}{\mathrm{wt}}

\def\F{\mathbbm{F}}

\def\dim{\mbox{dim}}
\def\Row{\text{Row}}
\def\rk{\mbox{rk}}

\def\1{\mathbbm{1}}

\DeclareUnicodeCharacter{00A0}{ }

\newtheorem{theo}{Theorem}

\newtheorem{lemma}[theo]{Lemma}

\newtheorem{theorem}{Theorem}

\begin{document}

\title{Small quantum Tanner codes from left--right Cayley complexes}

\author{Anthony Leverrier}
\affiliation{Inria Paris, France}
\orcid{0000-0002-6707-1458}
\email{anthony.leverrier@inria.fr}

\author{Wouter Rozendaal}
\affiliation{IMB, UMR 5251, Université de Bordeaux, France}
\orcid{0009-0004-8290-6136}
\email{wouter.rozendaal@math.u-bordeaux.fr}

\author{Gilles Z\'emor}
\affiliation{IMB, UMR 5251, Université de Bordeaux, France}
\affiliation{Institut Universitaire de France}
\orcid{0000-0002-6041-9554}
\email{gilles.zemor@math.u-bordeaux.fr}

\maketitle

\begin{abstract}
Quantum Tanner codes are a class of quantum low-density parity-check codes that provably display a linear minimum distance and a constant encoding rate in the asymptotic limit. 
When built from left--right Cayley complexes, they can be described through a lifting procedure and a base code,  which we characterize. We also compute the dimension of quantum Tanner codes when the right degree of the complex is 2.
Finally, we perform an extensive search over small groups and identify instances of quantum Tanner codes with parameters $[[144,12,11]]$, $[[432,20,\leq 22]]$ and $[[576,28,\leq 24]]$ for generators of weight 9.
\end{abstract}

Recent convincing demonstrations of quantum error correction suggest that building a large-scale quantum computer may be feasible in the coming years~\cite{goo25, PNH25}. An important remaining obstacle is the sheer cost of quantum fault tolerance. To help reduce this overhead as much as possible, it is crucial to design better quantum codes that combine a large encoding rate with a good distance. For hardware architectures compatible with long-range interactions, such as ions~\cite{MK13}, neutral atoms~\cite{SWM10,HBS20}, photons~\cite{OFV09}, or even superconducting qubits~\cite{BCG24}, quantum low-density parity-check (LDPC) codes offer an attractive solution, with asymptotically good parameters~\cite{PK22, LZ22, DHL23}.  Given the rapid progress of quantum hardware, a pressing question is to understand whether quantum LDPC codes can be competitive at scales of a few hundred physical qubits which are directly relevant for realistic architectures.\\

In the moderate-size regime, beating the parameters $[[n,2,\sqrt{n}]]$ of the rotated toric code~\cite{kit03} that encodes 2 logical qubits in $n$ physical qubits with a distance $\sqrt{n}$ remains challenging. 
The family of bivariate bicycle (BB) codes~\cite{KP13,BCG24,LLS25} is especially impressive: for instance, the gross code admits parameters $[[144,12,12]]$ and could lead to a massive reduction of the overhead~\cite{YSR25}. 
More generally, two-block group algebra codes~\cite{LP24} and lifted product codes~\cite{PK21,SHR24} display excellent parameters at finite length. 
It is notable that these various constructions, starting from the bicycle codes of MacKay et al.~\cite{MMM06}, can be described by a lifting procedure, as explained by Panteleev and Kalachev~\cite{PK21}.\\

Although very general classes of quantum Tanner codes exist~\cite{MRL25,MRL25b,GZ25}, we focus here on those instantiated on a left--right Cayley complex~\cite{DEL22}, as introduced in~\cite{LZ22}. While this construction yields asymptotically good codes, their finite-size performance remains poorly understood at the moment. A recent study~\cite{RBD25} found promising explicit codes but significantly behind BB codes of the same size. We note that quantum Tanner codes defined on coverings of 2-dimensional geometrical complexes are also known to achieve good distance~\cite{GZ25} for about 100 qubits.\\

Here, we revisit the original construction of~\cite{LZ22} with an emphasis on the lifting point of view: a quantum Tanner code is obtained from small local codes and two lifts associated with multisets $A, B$ of some finite group $\G$. Concretely, two of the local codes are associated with $A$, and the other two with $B$. In general, computing directly the parameters of the quantum code from its local structure appears hopeless, but special cases are tractable. For instance, Theorem \ref{thm} gives the dimension of the quantum code when the set $B$ has size two and the corresponding local code is the repetition code. Beyond this case, it is possible to search over the local codes and lifts and directly compute the parameters of the resulting quantum Tanner codes.  
Our main findings are presented in Table \ref{table} and show some quantum Tanner codes obtained through an extensive (but not exhaustive) search. For this search, we focused on local codes corresponding to the repetition code $[2,1,2]$, the shortened $[6,3,3]$ and the extended $[8,4,4]$ Hamming codes.

\section{Quantum Tanner codes as lifts}

A quantum Tanner code is a CSS code~\cite{CS96,ste96b} defined by associating qubits with the squares of a square-complex, and defining generators at the vertices of the complex. In the original construction of~\cite{LZ22}, the complex is chosen to be a left--right Cayley complex associated with a group $\G$ and two ordered multisets  $A = (a_1, \ldots, a_{n_A}), B = (b_1, \ldots, b_{n_B})$ of $\G$ of size $n_A, n_B$. The supports of the generators are codewords of some local product codes defined on $\F_2^{n_A} \times \F_2^{n_B}$. 
While the square-complex description is convenient to prove lower bounds on the distance, the more algebraic description given below is also useful to construct explicit small instances. \\

The alternative description of a quantum Tanner code interprets it as the lift of a specific base code. 
While the lifting procedure is easy to use in the classical case to obtain families of LDPC codes, it is more delicate to apply in the quantum case. Only recently was a general lifting procedure defined for general CSS codes~\cite{gue25}. In the case of the quantum Tanner codes, however, the lifting is simplified by the specific structure of the left--right Cayley complex. \\

Let $C_0,C_1\subseteq \F_2^{n_A}$ and $C_0',C_1'\subseteq \F_2^{n_B}$ be binary linear codes with parity-check matrices $H_0,H_1$ and $H_0',H_1'$, so that $C_i=\ker H_i$ and $C_i'=\ker H_i'$.  
We also fix generator matrices $G_0,G_1$ and $G_0',G_1'$ satisfying $\ker G_i = C_i^\perp$ and $\ker G_i' = {C_i'}^\perp$ (equivalently, the rows of $G_i$ span $C_i$, and likewise for $G_i'$).
We first define an (unlifted) \emph{base CSS code} on $n_A n_B$ qubits (arranged as an $n_A\times n_B$ grid) by
\begin{equation}\label{eq:base}
H_X^{\mathrm{base}}=
\begin{pmatrix}
H_0\otimes G_0'\\
H_1\otimes G_1'
\end{pmatrix},
\qquad
H_Z^{\mathrm{base}}=
\begin{pmatrix}
G_0\otimes H_1'\\
G_1\otimes H_0'
\end{pmatrix}.
\end{equation}
We will view this base code as the “local template’’ that is lifted by a group action. It is immediate that $H^{\text{base}}_X\cdot (H^{\text{base}}_Z)^{\mathsf T}=0$ showing that this is a valid CSS code. \\

Let
\[
C_{01}:=C_0\cap C_1,\qquad C_{01}^\perp:=C_0^\perp\cap C_1^\perp,
\qquad
C'_{01}:=C'_0\cap C'_1,\qquad C'^\perp_{01}:=C'^\perp_0 \cap C'^\perp_1.
\]
Note that $C_{01}^\perp$ denotes the intersection of the dual codes, and not the dual of $C_{01}$. 
Write
$k_{01}:=\dim \, C_{01}$, $k^\perp_{01}:=\dim \, C_{01}^\perp$, and similarly $k'_{01}$, $k'^\perp_{01}$ on the primed side.
Let $d_{01},d^\perp_{01},d_{01}',d'^\perp_{01}$ denote the minimum distance of these four intersection codes. By convention, the distance of the 0 code is infinite.

\begin{lemma}[Base-code parameters]\label{lem}
The base code~\eqref{eq:base} has length $n=n_A n_B$ and dimension
\[
k \;=\; k_{01}\,k_{01}' \;+\; k_{01}^\perp\, k'^\perp_{01}.
\]
Moreover, there exists a symplectic basis of logical operators in which every logical $X$-operator or logical $Z$-operator is supported on a single row or on a single column of the $n_A\times n_B$ qubit array. When $k>0$, the minimum distance of the base code is 
\[ d=\min(d_{01},d_{01}',d_{01}^\perp,d'^\perp_{01}).\]
\end{lemma}

The lemma is proved in Appendix \ref{app}.
Many natural choices of local codes are such that $\mathrm{dim}(C_0) + \mathrm{dim}(C_1) = n_A$ and $\mathrm{dim}(C_0') + \mathrm{dim}(C_1') = n_B$, for instance codes with parameters $[2,1,2], [6,3,3]$ or $[8,4,4]$. The dimension of the base quantum Tanner code then simplifies to
\begin{align*}
k = 2\, k_{01}\,k_{01}'.
\end{align*}

Quantum Tanner codes are built by lifting this base structure along a finite group $\G$.
We fix multisets $A=(a_1,\dots,a_{n_A})$ and $B=(b_1,\dots,b_{n_B})$ of elements of $\G$.
The lift replaces each base qubit indexed by $(i,j)\in[n_A]\times[n_B]$ with a fiber of size $|\G|$, indexed by $g\in \G$, so the lifted code has
\[
n = n_A n_B |\G|
\]
qubits in total. We use the left and right regular actions of $\G$: for $a,b\in \G$, let $\lambda_a,\rho_b\in\F_2^{|\G|\times |\G|}$ be the permutation matrices defined by
\[
(\lambda_a)_{g,h}=\delta_{h,ag},
\qquad
(\rho_b)_{g,h}=\delta_{h,gb^{-1}} .
\]
These actions commute: $\lambda_a\rho_b=\rho_b\lambda_a$ for all $a,b\in \G$.

We define block-diagonal permutations acting on the lifted qubits by
\[
L_A := \bigoplus_{i=1}^{n_A}\ \bigoplus_{j=1}^{n_B}\ \lambda_{a_i},
\qquad
R_B := \bigoplus_{i=1}^{n_A}\ \bigoplus_{j=1}^{n_B}\ \rho_{b_j}.
\]
With these notations, the lifted quantum Tanner code has check matrices
\begin{align}\label{eq:lifted}
H_X=
\begin{pmatrix}
H_0\otimes G_0'\otimes I_{|\G|}\\[2pt]
\bigl(H_1\otimes G_1'\otimes I_{|\G|}\bigr)\,L_A R_B
\end{pmatrix},
\qquad
H_Z=
\begin{pmatrix}
\bigl(G_0\otimes H_1'\otimes I_{|\G|}\bigr)\,R_B\\[2pt]
\bigl(G_1\otimes H_0'\otimes I_{|\G|}\bigr)\,L_A
\end{pmatrix}.
\end{align}
We index qubits by triples $(i,j,g) \in [n_A] \times [n_B]\times \G$. The permutation $L_A$ acts as $(i,j,g) \mapsto (i,j,a_i g)$ and the permutation $R_B$ acts as $(i,j,g) \mapsto (i,j,g b_j^{-1})$.

\begin{figure}[H]
  \centering
  \includegraphics[width=0.73\columnwidth]{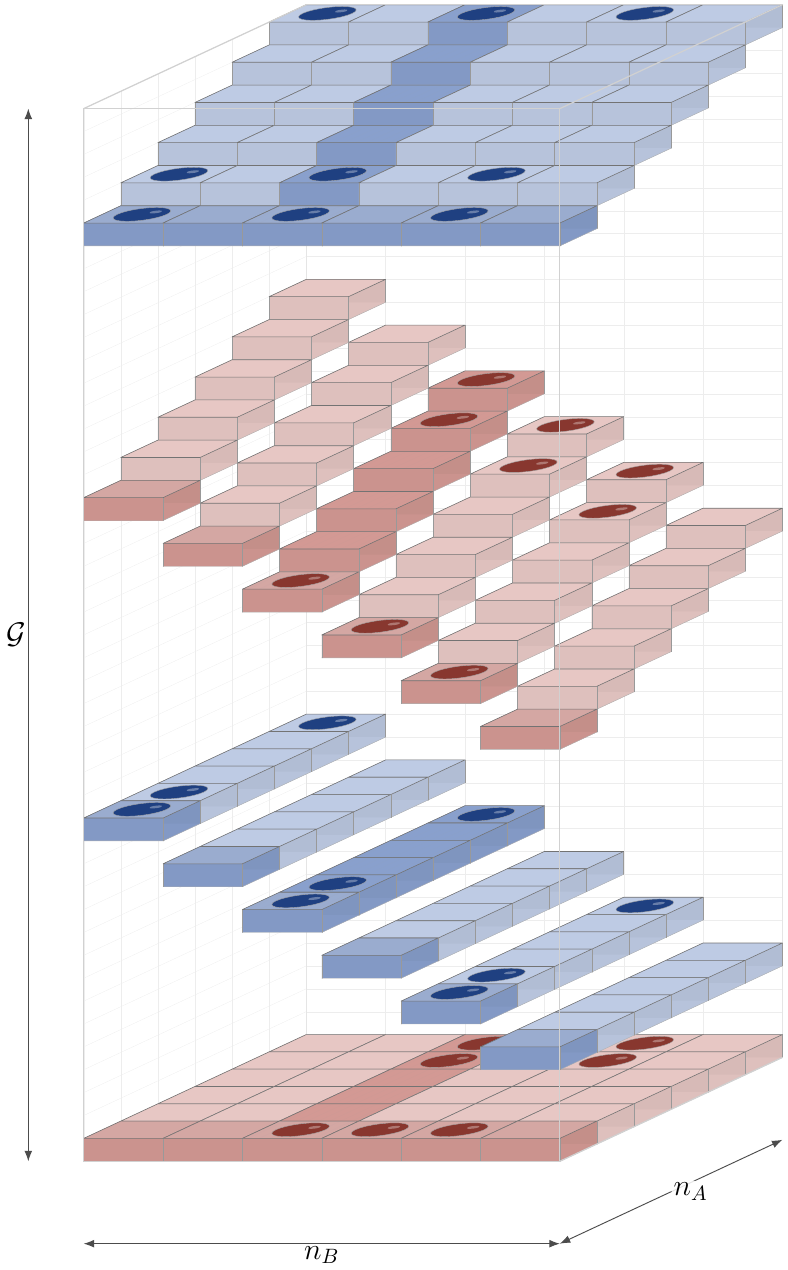}
\caption{Three-dimensional representation of the quantum Tanner codes, with $\G = C_{46}, n_A=n_B=6$ and $A = (0,1,2,3,4,5), B=(0,2,4,6,8,10)$, where we use additive notation for cyclic groups.
Qubits are associated with tiles indexed by $(i,j,g)$ with $i \in[n_A], j \in [n_B], g\in \G$.
Each generator is supported within the translate of a horizontal slice. The figure depicts 4 generators, each associated with specific choices of local codewords (marked with dark disks) and a group element. The whole set of generators is obtained by considering a basis of the local product codes, and all $\G$-translates.
From bottom to top, an $X$-generator defined on the (red) set $\{ (i,j, g)  :  i \in [n_A], j\in [n_B]\}$ corresponding to a row of $H_0 \otimes G_0' \otimes I_{|\G|}$, a $Z$-generator defined on the (blue) set $\{ (i,j, gb_j^{-1})  :  i \in [n_A], j\in [n_B]\}$  corresponding to a row of $(G_0 \otimes H_1' \otimes  I_{|\G|})R_B$, an $X$-generator defined on the set $\{ (i,j, a_i gb_j^{-1})  :  i \in [n_A], j\in [n_B]\}$ corresponding to a row of $(H_1 \otimes G_1' \otimes I_{|\G|})L_A R_B$, and a $Z$-generator defined on the set $\{ (i,j, a_i g)  :  i \in [n_A], j\in [n_B]\}$  corresponding to a row of $(G_1 \otimes H_0' \otimes  I_{|\G|})L_A$.
The support of an $X$-generator can only overlap that of a $Z$-generator on a single row or column of such  slices (depicted with a darker color on the figure): in that case, the supports correspond to orthogonal codewords of the local codes, ensuring the commutation of the generators.
}
  \label{fig:tanner3d}
\end{figure}

Each stabilizer generator of the lifted code is supported  within a single “slice’’ indexed by $(i,j)\in[n_A]\times[n_B]$, together with a translate in the $\G$-fiber.  
More precisely, the $X$-type generators correspond to product codewords in
$C_i^\perp\otimes C_i'$ (for $i=0,1$), while the $Z$-type generators correspond to product codewords in
$C_0\otimes {C_1'}^\perp$ and $C_1\otimes {C_0'}^\perp$, matching the block structure of~\eqref{eq:lifted}.  
This slice structure is naturally encoded by the left--right Cayley complex associated with $(A,B,\G)$, which we illustrate in Figure~\ref{fig:tanner3d}.
The support of an $X$-generator can only overlap that of a $Z$-generator on a single row or column of such  slices (depicted with a darker color on the figure): in that case, the supports correspond to orthogonal codewords of the local codes, ensuring the commutation of the generators.\\

Computing the dimension of the quantum Tanner codes of \eqref{eq:lifted} is more delicate than for the base code: the lift can introduce additional linear relations among checks, or remove them, and $k$ may differ from the base value.  This is not surprising since it is already the case for classical Tanner codes, or more generally for classical LDPC codes constructed from protographs~\cite{tho03}.
In the special case $n_B=2$ with a repetition local code on the $B$-side, we observe a particularly simple dimension behavior (proved in Appendix \ref{app}).

\begin{theorem}\label{thm}
Assume $|B|=2$ and that the $B$-side local codes are the length-$2$ repetition code, $H_0'=H_1'= (1 \quad 1)$.
Write $B=(b_1,b_2)$.  Assume moreover that right multiplication by $b_1^{-1}b_2$ acts transitively on $\G$.
Then the lifted quantum Tanner code~\eqref{eq:lifted} has dimension
\[
k = k_{01} + k_{01}^\perp,
\]
i.e., the same dimension as the base code.
\end{theorem}

The distance of quantum Tanner codes is even more challenging to predict. A nice special case is for odd $|\G|$ when the dimensions of the base code and lifted code coincide. Then the operators obtained by taking $|\G|$ copies of the logical operators of the base code are again nontrivial logical operators in the lifted code. The distance of the lifted code is therefore at least the distance of the base code and at most $|\G|$ times larger. These observations were already made in~\cite{GZ25} and~\cite{SRB25}. Obviously, the point of lifting the base code is to significantly increase the distance, and this bound is too loose to be very useful.\\

Restricting the set of qubits to a single vertical $A$-slice $\{(i,j_*,g)  :  i \in [n_A], g\in \G\}$ or $B$-slice $\{({i_*}, j,g)  :  j \in [n_B], g\in \G\}$, one can define classical codes with parity-check matrices 
\[
\begin{pmatrix}H_0\otimes  I_{|\G|}\\ (H_1\otimes I_{|\G|})L_A\end{pmatrix},
\quad
\begin{pmatrix}G_0\otimes I_{|\G|}\\ (G_1\otimes I_{|\G|})L_A\end{pmatrix}
\]
for $A$-slices, and
\[
\begin{pmatrix}G_0'\otimes I_{|\G|}\\ (G_1'\otimes I_{|\G|})R_B\end{pmatrix},
\quad
\begin{pmatrix}H_0'\otimes I_{|\G|}\\ (H_1'\otimes I_{|\G|})R_B\end{pmatrix}
\]
for $B$-slices. These are classical Tanner codes~\cite{T81} defined on the Cayley graph $\mathrm{Cay}_2(\G,A)$ with local codes $C_i, C^\perp_i$, and on $\mathrm{Cay}_2(\G,B)$ with local codes $C'_i$ and $C'^\perp_i$. The graphs are defined as bipartite graphs with two sets of vertices $V_0 = \G \times \{0\}, V_1 = \G \times \{1\}$ and edges given respectively by $(g,0) \sim (ag,1)$ and $(g,0) \sim (gb^{-1},1)$.
Codewords of these classical Tanner codes automatically satisfy the relevant constraints of \eqref{eq:lifted}. Any classical codeword of the $A$-type Tanner code therefore gives rise to a logical $Z$-operator supported on a single vertical slice and any classical codeword of the $B$-type code yields a logical $X$-operator for the quantum Tanner code. Note that these are not the only logical operators of the quantum code, and moreover, they are not necessarily nontrivial logical operators. 
\\

\section{Numerical search}

Since it seems implausible that one can compute the parameters of the quantum Tanner code directly from the local structure and choice of the lift, it makes sense to search for good codes by trying many choices of local codes and multisets $A$ and $B$. 
Computing the dimension of the resulting code can be done exactly by Gaussian elimination. There does not exist any efficient method for evaluating the distance, as the best known deterministic algorithms display exponential complexity. Fortunately, for moderate-size codes of a few hundred physical qubits, probabilistic algorithms provide reliable estimates. We used the GAP package \texttt{QDistRnd}~\cite{QDistRnd,PSK22}, which returns an upper bound for the distance. While the results are never guaranteed to give the true value of the distance, we can increase our confidence by considering a large number of trials. In the tables, we report the best value found by \texttt{QDistRnd} after the stated number of trials (50 000 trials unless specified otherwise).

The crucial elements of the construction are the local codes $C_i, C_i'$. Recall that the support of the generators coincide with a product of two codewords of these local codes. In particular, this implies that the weight $w$ of the generators is bounded by
\[ w \geq \max (d_0^\perp d'_0, d^\perp_1 d'_1, d_0 d'^\perp_1, d_1 d'^\perp_0),\]
where $d_i, d_i^\perp$  are the distances of $C_i$ and $C^\perp_i$, and similarly for the primed codes.
We seek to minimize this value for hardware implementations, and focus here on local codes that give reasonably small weights, namely $C_i$ with parameters $[6,3,3]$ or $[8,4,4]$, and $C_i'$ with parameters $[2,1,2]$ or $[6,3,3]$. All these codes have dual codes with identical parameters. The various combinations yield quantum Tanner codes with generators of weight 6, 8 or 9.
One could use only repetition codes $[2,1,2]$, thus getting quantum Tanner codes with generators of weight 4, but these would not beat the (rotated) toric code which is obtained in the special case where $\G$ is the product of two cyclic groups of the same order, and $A, B$ consist of generators of the component subgroups. \\

For the sets of parameters $[6,3,3]$ and $[8,4,4]$ corresponding to the shortened and extended Hamming codes, there are 30 distinct possible codes that are obtained from the canonical parity-check matrices
\begin{align*}
H_{[6,3,3]} &=
\begin{pmatrix}
1 & 0 & 0 & \vline & 0 & 1 & 1 \\
0 & 1 & 0 & \vline & 1 & 0 & 1 \\
0 & 0 & 1 & \vline & 1 & 1 & 0
\end{pmatrix}
\qquad &
G_{[6,3,3]} &=
\begin{pmatrix}
0 & 1 & 1 & \vline & 1 & 0 & 0 \\
1 & 0 & 1 & \vline & 0 & 1 & 0 \\
1 & 1 & 0 & \vline & 0 & 0 & 1
\end{pmatrix},
\\[0.8em]
H_{[8,4,4]} &=
\begin{pmatrix}
1 & 0 & 0 & 0 & \vline & 0 & 1 & 1 & 1 \\
0 & 1 & 0 & 0 & \vline & 1 & 0 & 1 & 1 \\
0 & 0 & 1 & 0 & \vline & 1 & 1 & 0 & 1 \\
0 & 0 & 0 & 1 & \vline & 1 & 1 & 1 & 0
\end{pmatrix}
\qquad &
G_{[8,4,4]} &=
\begin{pmatrix}
0 & 1 & 1 & 1 & \vline & 1 & 0 & 0 & 0 \\
1 & 0 & 1 & 1 & \vline & 0 & 1 & 0 & 0 \\
1 & 1 & 0 & 1 & \vline & 0 & 0 & 1 & 0 \\
1 & 1 & 1 & 0 & \vline & 0 & 0 & 0 & 1
\end{pmatrix}.
\end{align*}
through a column permutation.
Since the parameters of the quantum Tanner code do not change under a global permutation of the qubits, it is in fact sufficient to pick $H_0$ or $H_0'$ of the form above, and only try the 30 permutations for $H_1$ and $H_1'$. The picture is much simplified when the local code has length 2 since the only possibility is $H=G = (1 \quad 1)$, so no permutation need be considered.\\

Similarly, for a group $\G$, it is not necessary to consider all possible multisets $A$ and $B$ of the appropriate size. It is sufficient to look at multisets that yield inequivalent Cayley graphs $\mathrm{Cay}_2(\G,A)$ and $\mathrm{Cay}_2(\G,B)$.\\

We conducted two main searches. We focused first on the case of Theorem \ref{thm} where $n_B=2$ and the local code is the repetition code. We considered cyclic groups $C_m$, so that the number of qubits equals either $12m$ or $16m$ and exhaustively looked at codes with lengths up to approximately 250. In that setting, the quantum Tanner code dimension coincides with the dimension of the base code. We found that codes of dimension 2 were the most interesting ones in general, and could give fairly large distances such that $d^2/n > 1$, as reported in Table \ref{tab:w6w8}.\\

The second search we performed was for the case where all the local codes (and their dual) have parameters $[6,3,3]$. The code length is $n=36 |\G|$. Given that the search space quickly becomes much larger, especially when we increase the group size, it is useful to only consider multisets $A$ and $B$ such that the classical Tanner codes described at the end of the previous section already display a large distance. These classical codes have length $6|\G|$ and estimating their distance is much quicker than estimating the distance of the full quantum Tanner code. While it is not always the case that a codeword of these codes will become a nontrivial logical operator for the quantum Tanner code (they are sometimes mapped to stabilizers), it turns out to be a very effective heuristic to filter out many potential bad choices of lifts. 
We report the results in Table \ref{tab:w9}. In particular, we found several instances where both $k$ and $d$ are fairly large, for example $\llbracket 576, 28, \geq 24\rrbracket$ that displays $k > \sqrt{n}$ and $d = \sqrt{n}$. \\

\begin{table}[t]
\centering
\begin{subtable}[t]{0.53\textwidth}
\centering
\renewcommand{\arraystretch}{1.05}
\begin{tabular}{@{}rrrr @{\qquad} rrrr@{}}
\toprule
\multicolumn{4}{c}{$w=6$} & \multicolumn{4}{c}{$w=8$} \\
\cmidrule(r){1-4}\cmidrule(l){5-8}
$n$ & $k$ & $d$ & $\frac{d^2}{n}$ & $n$ & $k$ & $d$ & $\frac{d^2}{n}$ \\
\midrule
  60 & 2 &  8 & 1.1 &   80 & 2 & 10 & 1.3 \\
  72 & 2 &  9 & 1.1 &  112 & 2 & 12 & 1.3 \\
  84 & 2 & 10 & 1.2 &  128 & 6 & 12 & 1.1 \\
 108 & 2 & 12 & 1.3 &  144 & 2 & 14 & 1.4 \\
 132 & 2 & 14 & 1.5 &  160 & 2 & 16 & 1.6 \\
 168 & 2 & 16 & 1.5 &  192 & 2 & 18 & 1.7 \\
 216 & 2 & 18 & 1.5 &  224 & 2 & 20 & 1.8 \\
 252 & 2 & 20 & 1.6 &  256 & 2 & 22 & 1.9 \\
\bottomrule
\end{tabular}
\caption{Best cyclic-group instances found when the $B$-code is the repetition code $[2,1,2]$ and the $A$-code has parameters $[6,3,3]$ or $[8,4,4]$, giving generator weight 6 and 8, respectively. Distances are estimated with QDistRnd~\cite{QDistRnd}, using either 50 000 trials, or 500 000 trials when the reported distance is at least 20.}
\label{tab:w6w8}
\end{subtable}
\hfill
\begin{subtable}[t]{0.43\textwidth}
\centering

\renewcommand{\arraystretch}{1.05}
\begin{tabular}{@{}rrr c c@{}}
\toprule
\multicolumn{5}{c}{$w=9$} \\
\midrule
$n$ & $k$ & $d$ & group & \# trials\\
\midrule
144 & 8  & 12 & $C_2 \times C_2$       & 50k \\
144 & 12 & 11 & $C_2 \times C_2$       & 50k \\
216 & 8  & 18 & $C_6$                  & 50k \\
288 & 16 & 16 & $Q_8$                  & 50k \\
288 & 8  & 19 & $C_8$                  & 50k \\
324 & 4  & 26 & $C_9$                  & 3M \\
396 & 2  & 29 & $C_{11}$               & 3M \\
432 & 20 & 22 & $C_6 \times C_2$        & 1M \\
432 & 24 & 18 & $C_6 \times C_2$        & 200k \\
576 & 28 & 24 & $C_4 \rtimes C_4$       & 3M \\
\bottomrule
\end{tabular}
\caption{Parameters of quantum Tanner codes for $[6,3,3]$ local codes and generator weight $9$.
The last column indicates the number of random trials used for the distance estimation~\cite{QDistRnd}.}
\label{tab:w9}
\end{subtable}

\caption{Best quantum Tanner codes found through numerical search. All parity-check matrices are available as auxiliary files on arXiv.}
\label{table}

\end{table}

\section{Discussion and open questions}

While quantum Tanner codes constructed from left--right Cayley complexes were known to be asymptotically good, it is instructive to see that they can be competitive at moderate size. Choosing all local codes to be shortened Hamming codes $[6,3,3]$ appears to give good quantum parameters. 
It is likely that much better codes exist if we allow $n$ to be between 500 and 1000. There are difficulties to explore that regime properly: first the search space becomes very large and even the heuristics mentioned above quickly become inefficient. Second, we are lacking fast algorithms to estimate the distance of quantum codes with parameters $n \geq 500, d \geq 20$.
One could try a more aggressive filtering that only keeps the lifts such that the classical Tanner codes living in vertical slices are close to optimal. This is indeed easy to do, but we observed that this misses the quantum codes with the best distances. \\

An interesting open question is to better understand the structure of the logical operators. When $|\G|$ is odd and the dimensions of the lifted and base codes coincide, then each logical operator of the base code yields a logical operator of the quantum Tanner code corresponding to $|\G|$ copies of the base logical. One expects that many logical operators indeed live in single vertical slices and correspond to codewords of the classical Tanner codes. But these are not the only operators, and we have checked that in general, it is not possible to find a basis of logical operators where each one has support in a single vertical slice. \\

Finally, it would be very interesting to understand how to decode short quantum Tanner codes. Efficient decoding algorithms exist~\cite{LZ25,LZ23} in the asymptotic regime, but their performance guarantees rely on robustness properties of the local codes that are not satisfied for small instances.

\begin{acknowledgements}
This work was funded by the Plan France 2030 through the project ANR-22-PETQ-0006.
\end{acknowledgements}

%\bibliographystyle{quantum}
%\bibliography{biblio}

\appendix

\section{Deferred proofs}
\label{app}

\subsection{Proof of Lemma \ref{lem}}

We write the base qubits as an $n_A\times n_B$ array, so that $\F_2^{n_A n_B}\cong \F_2^{n_A}\otimes \F_2^{n_B}$.
Recall that the $X$-check space is
\[
\mathcal X := \Row(H_X^{\mathrm{base}}) \;=\; (C_0^\perp\otimes C_0') \;+\; (C_1^\perp\otimes C_1')
\]
and the $Z$-check space is
\[
\mathcal Z := \Row(H_Z^{\mathrm{base}}) \;=\; (C_0\otimes C_1'^\perp) \;+\; (C_1\otimes C_0'^\perp).
\]
We denote by $k_i, k_i'$ the dimension of $C_i$ and $C_i'$. 
\paragraph{Dimension.}
Since $\dim(U\otimes V)=\dim(U)\, \dim(V)$, we have
\[
\dim(C_0^\perp\otimes C_0')=(n_A-k_0)k_0',
\qquad
\dim(C_1^\perp\otimes C_1')=(n_A-k_1)k_1'.
\]
Moreover, one checks that
\[
(C_0^\perp\otimes C_0')\cap(C_1^\perp\otimes C_1')
=(C_0^\perp\cap C_1^\perp)\otimes(C_0'\cap C_1').
\]
Therefore
\begin{align*}
\rk(H_X^{\mathrm{base}})&=\dim \, \mathcal X
=(n_A-k_0)k_0' + (n_A-k_1)k'_1 - k_{01}^\perp k'_{01}\\
&=  n_A (k_0' + k_1') - k_0 k_0' - k_1 k_1' - k_{01}^\perp k_{01}'
\end{align*}
Similarly,
\[
(C_0\otimes C_1'^\perp)\cap(C_1\otimes C_0'^\perp)
=(C_0\cap C_1)\otimes(C_0'^\perp\cap C_1'^\perp),
\]
so
\begin{align*}
\rk(H_Z^{\mathrm{base}})&=\dim \, \mathcal Z
=k_0(n_B-k_1') + k_1(n_B-k_0') - k_{01} k'^\perp_{01}\\
&= n_B (k_0 + k_1) - k_0 k_1' - k_1 k_0' - k_{01} k'^\perp_{01}
\end{align*}
Recalling that $C_0^\perp \cap C_1^\perp = (C_0 + C_1)^\perp$, we have
\begin{align*}
k^\perp_{01} &= n_A - \mathrm{dim} (C_0+C_1) \\
 &= n_A - k_0 - k_1 + \mathrm{dim}(C_0 \cap C_1) \\
 &= n_A - k_0 - k_1 + k_{01}
 \end{align*}
 and similarly $n_B - k'_0 - k'_1 = k'^\perp_{01} - k'_{01}$.\\
Finally, the CSS dimension formula gives
\begin{align*}
k&= n_An_B -  \mathrm{rk} \, H^{\text{base}}_X - \mathrm{rk} \, H^{\text{base}}_Z\\
&= n_An_B- n_A (k_0'+k_1') - n_B (k_0+k_1) + (k_0+k_1)(k_0'+k_1') + k_{01}^\perp k_{01}'+k_{01} k'^\perp_{01}\\
&= (n_A -k_0-k_1)(n_B- k'_0-k'_1) + k_{01}^\perp k_{01}'+k_{01} k'^\perp_{01}\\
&= (k^\perp_{01} - k_{01}) (k'^{\perp}_{01} - k'_{01}) + k_{01}^\perp k_{01}'+k_{01} k'^\perp_{01}\\
&= k_{01} k'_{01} + k^\perp_{01} k'^\perp_{01}.
\end{align*}

\paragraph{Row/column logical basis.} The argument is similar to the case of hypergraph product codes~\cite{TZ14} studied in~\cite{QWV23}.
Choose bases $\{u_1,\dots,u_{k_{01}}\}$ of $C_{01} = C_0\cap C_1$ and $\{v_1,\dots,v_{k_{01}'}\}$ of $C_{01}' = C'_0 \cap C'_1$ in reduced row-echelon form.
In particular, there exist pivot index sets $I=\{i_1,\dots,i_{k_{01}}\}\subseteq[n_A]$ and $J=\{j_1,\dots,j_{k_{01}'}\}\subseteq[n_B]$ such that
\[
u_p(i_{p'})=\delta_{p,p'},
\qquad
v_q(j_{q'})=\delta_{q,q'}.
\]
Define, for $p\in[k_{01}]$ and $q\in[k_{01}']$,
\[
\bar Z_{p,q} := u_p\otimes e_{j_q}\in \F_2^{n_A}\otimes\F_2^{n_B},
\qquad
\bar X_{p,q} := e_{i_p}\otimes v_q\in \F_2^{n_A}\otimes\F_2^{n_B},
\]
where $e_j$ denotes the standard basis vector.
Then $\bar Z_{p,q}\in \ker H_X^{\mathrm{base}}$ because $u_p\in C_0\cap C_1$, and
$\bar X_{p,q}\in \ker H_Z^{\mathrm{base}}$ because $v_q\in C_0'\cap C_1'$.
Moreover,
\[
\langle \bar X_{p,q}, \bar Z_{p',q'}\rangle
= u_{p'}(i_p)\,v_q(j_{q'})=\delta_{p,p'}\delta_{q,q'},
\]
so these operators provide $k_{01}k_{01}'$ logical qubits with $X$-operators supported on a single row and $Z$-operators supported on a single column.

For the second summand in $k$, choose pivot bases
$\{u_1^\perp,\dots,u_{k_{01}^\perp}^\perp\}$ of $C_{01}^\perp=C_0^\perp\cap C_1^\perp$ with pivot indices $I^\perp=\{i_1^\perp,\dots,i_{k_{01}^\perp}^\perp\}\subseteq[n_A]$, and
$\{v^\perp_1,\dots,v^\perp_{k'^\perp_{01}}\}$ of ${C_{01}'}^\perp={C_0'}^\perp\cap {C_1'}^\perp$ with pivot indices $J^\perp=\{j_1^\perp,\dots,j_{{k_{01}'}^\perp}^\perp\}\subseteq[n_B]$.
Define
\[
\bar X^\perp_{r,s} := u_r^\perp\otimes e_{j_s^\perp},
\qquad
\bar Z^\perp_{r,s} := e_{i_r^\perp}\otimes v^\perp_s.
\]
Then $\bar X^\perp_{r,s}\in\ker H_Z^{\mathrm{base}}$ and $\bar Z^\perp_{r,s}\in\ker H_X^{\mathrm{base}}$, and the same pivot argument shows that these form $k_{01}^\perp k'^\perp_{01}$ additional logical qubits, again supported on single rows/columns. These logical qubits are different from the first $k_{01}k_{01}'$ qubits since $\bar X_{p,q}$ commute with $\bar Z^\perp_{r,s} $ because $v_q$ is orthogonal to $v^\perp_s$, and $\bar Z_{p,q}$ commute with $\bar X^\perp_{r,s}$ because $u_p$ is orthogonal to $u^\perp_r$.
Altogether we obtain a symplectic basis of logical operators with the claimed support property.

\paragraph{Distance.}
The row and column operators described above give the upper bound on the distance.

Let us turn to the lower bound on the distance $d_Z$, corresponding to the minimal weight of an element of $\ker H^{\text{base}}_X \setminus  \mathrm{Im} (H^{\text{base}}_Z)^{\mathsf T}$ and consider a nontrivial $Z$-logical operator $M \in \ker H^{\text{base}}_X$. 
First note that this implies that 
\[ H_0 MG_0'^{\mathsf T} = 0, \qquad H_1 M G_1'^{\mathsf T} = 0.\]
In particular, for any basis vector $v_q \in C'_0 \cap C'_1$, it holds that $H_0 M v_q^{\mathsf T} = 0$ and $H_1 M v_q^{\mathsf T} =0$, which means that $M v_q^{\mathsf T} \in C_{01}$. The vector $M v_q^{\mathsf T}$ is unchanged if we add a $Z$-stabilizer to $M$, hence $M v_q^{\mathsf T}$ only depends on the logical class of $M$.
Similarly, for a basis vector $u_r^\perp\in C_{01}^\perp$, we have $u_r^\perp M\in C'^\perp_{01}$ and also depends only on the logical class.

By construction of the logical basis, a nontrivial logical class has either a nonzero component along the $\bar Z_{p,q}$ family or a nonzero component along the $\bar Z_{r,s}^\perp$ family.
In the first case, there exists $q$ such that $M v_q^{\mathsf T}\neq 0$, hence $\wt(M)\ge \wt(M v_q^{\mathsf T})\ge d_{01}$.
In the second case, there exists $r$ such that $u_r^\perp M\neq 0$, hence $\wt(M)\ge \wt(u_r^\perp M)\ge d'^\perp_{01}$.
This proves the claimed formula for $d_Z$.
The argument for $d_X$ is identical using the $X$-logical basis.
Finally $d=\min(d_X,d_Z)$ yields the distance statement in Lemma~\ref{lem}.
\qed

\subsection{Proof of Theorem \ref{thm}}

Assume $|B|=2$ with $B=(b_1,b_2)$ and $C_0'=C_1'$ to be the 2-repetition code, so that $G_0'=G_1'=(1\ 1)$ and $H_0'=H_1'=(1\ 1)$.
Group the qubits as two vertical slices (corresponding to the two elements of $B$), each slice containing $n_A|\G|$ qubits.

Let $Q$ denote the permutation on $\G$ induced by right multiplication by $b_1^{-1}b_2$.
By hypothesis, $Q$ is a single cycle, hence the fixed subspace $\{x\in\F_2^{|\G|} : xQ=x\}$ is one-dimensional, spanned by the all-ones vector $\mathbf 1_G$.

The parity-check matrix $H_X$ is a vertical concatenation of two blocks, corresponding to the two blocks in~\eqref{eq:lifted}.
Each block has rank $(n_A-k_i)|\G|$ (since $G_0'=G_1'$ has rank $1$ and $L_AR_B$ is a permutation), so
\[
\rk(H_X)=(n_A-k_0)|\G|+(n_A-k_1)|\G|-\dim\bigl(\Row(X_0)\cap \Row(X_1)\bigr),
\]
where $X_0:=H_0\otimes (1\ 1)\otimes I_{|\G|}$ and $X_1:=(H_1\otimes (1\ 1)\otimes I_{|\G|})L_AR_B$.

A row vector lies in $\Row(X_0)$ iff its two vertical slices are identical and, for each $g\in \G$, the induced length-$n_A$ vector belongs to $C^\perp_0$.
A row vector lies in $\Row(X_1)$ iff its two slices become identical after undoing the $R_B$ permutation; equivalently, the two layers are related by the permutation $Q$ on the $\G$-fiber, and the induced length-$n_A$ vectors lie in $C_1^\perp$.
Therefore, an element of $\Row(X_0)\cap\Row(X_1)$ corresponds precisely to a choice of $x\in \F_2^{n_A}$ such that $x\in C_0^\perp\cap C_1^\perp$ and the $\G$-fiber is constant in each $A$-block.
Hence
\[
\dim\bigl(\Row(X_0)\cap \Row(X_1)\bigr)=\dim(C_0^\perp\cap C_1^\perp)=k_{01}^\perp,
\]
and we conclude
\[
\rk(H_X)=(2n_A-k_0-k_1)|\G|-k_{01}^\perp.
\]

The same argument applies to $H_Z$, whose two blocks have ranks $k_0|\G|$ and $k_1|\G|$.
Now $\Row(Z_1)$ (the block multiplied by $L_A$) enforces equality of the two slices, while $\Row(Z_0)$ (the block multiplied by $R_B$) relates them by $Q$. One obtains as before
\[
\dim\bigl(\Row(Z_0)\cap \Row(Z_1)\bigr)=\dim(C_0\cap C_1)=k_{01},
\]
hence
\[
\rk(H_Z)=(k_0+k_1)|\G|-k_{01}.
\]

Since the lifted code has $n=2n_A|\G|$ qubits, the CSS dimension formula gives
\[
k=n-\rk(H_X)-\rk(H_Z)
=2n_A|\G|-\bigl((2n_A-k_0-k_1)|\G|-k_{01}^\perp\bigr)-\bigl((k_0+k_1)|\G|-k_{01}\bigr)
= k_{01}+k_{01}^\perp.
\]
This proves Theorem~\ref{thm}.
\qed

\end{document}